\documentclass{iise}

\usepackage{multirow}
\usepackage[utf8]{inputenc}

\usepackage{amsmath,amssymb,amsthm}
\usepackage{mathtools}
\usepackage{dsfont}
\usepackage{wrapfig}

\usepackage{xcolor}
\usepackage{color}
\usepackage[normalem]{ulem}

\usepackage{hyperref}

\usepackage{algorithm}
\usepackage{algorithmic}

\usepackage{tikz}

\usepackage{graphicx}
\usepackage{caption}
\usepackage{subcaption}
\usepackage{float}

\usepackage{booktabs}
\usepackage{enumitem}

\conference{Proceedings of the IISE Annual Conference \& Expo 2026\\
Y. Xiang, D., Yu, R. Thiesing, eds.}

\title{\titlesize Optimal Reconfiguration of Distributed Battery Networks Under Connectivity and Energy Constraints}

\authorlist{KC$^*$, Taghieh, Palacios, Rashidioun, Swissler, and Park}

\abstractID{15835}

\begin{document}
\author{Pranay KC$^*$, Amin Taghieh, Maria Angel Palacios, Mohammadali Rashidioun, \\ Petras Swissler, and SangWoo Park\\
\textit{New Jersey Institute of Technology, Newark, NJ}\\
Contact: prk5@njit.edu, sangwoo.park@njit.edu}

\maketitle

\setlength{\bibsep}{1pt}
\setlength{\floatsep}{4pt}
\setlength{\textfloatsep}{4pt}

\titlespacing{\section}{0pt}{10pt}{-4pt}
\titlespacing{\subsection}{0pt}{10pt}{-4pt}

\begin{abstract}
Networked battery systems arise in industrial automation, distributed energy applications, and multi-agent systems, where terminals consume energy locally and recharge only when connected to a source. Resource constraints often limit the number of simultaneous connections, requiring networks to be dynamically reconfigured to maintain system functionality. Managing such networks in dynamic environments is challenging, particularly when low-energy terminals must be prioritized for timely replenishment.
This paper presents a battery-aware topology optimization algorithm that extends the GeoSteiner framework with a tailored Mixed-Integer Linear Program (MILP) formulation for Full Steiner Tree (FST) aggregation. The formulation minimizes network length while prioritizing low-battery terminals through a weighted objective subject to a global budget constraint, enabling partial network formation under realistic resource limits. An overlap-correction term is introduced that prevents double-counting when selected trees share terminals. To capture the network reconfiguration cost between time steps, a graph-distance metric penalizes frequent topology changes, resulting in 72.2\% reduction compared to a baseline without penalty.
Simulations on a 20-terminal network demonstrate battery levels are effectively managed as the lowest battery level improved from 2.7\% to 68.6\% over 30 iterations while maintaining the topology stability and budget utilization (92\%). The framework offers a principled approach to designing energy-aware, adaptive connectivity in power-limited multi-agent systems.
\end{abstract}

\section*{Keywords}
Topology optimization, Steiner tree, mixed-integer linear programming, battery management, swarm robotics

\vspace{-2mm}
\section{Introduction} \label{sect:introduction}

\subsection{Background and Motivation}

Networked battery systems arise in industrial automation, distributed energy applications, and multi-agent systems, where terminals consume energy locally and recharge only when connected to a source. Resource constraints often limit the number of simultaneous connections, requiring networks to be dynamically reconfigured to maintain system functionality. Managing such networks in dynamic environments is challenging, particularly when low-energy terminals must be prioritized for timely replenishment.

The problem of designing optimal network structures to address connectivity and resource challenges has been extensively studied under the framework of topology optimization. A common approach is to formulate topology control as a mathematical optimization problem, typically using MILP or metaheuristic techniques such as simulated annealing. In wireless sensor networks, Thammawichai and Luangwilai~[1] formulated topology control as a MILP to minimize energy consumption and maximize network lifetime, using clustering and sleep/wake-up schemes to balance load across nodes. In mobile optical wireless networks, Dwivedi~\textit{et al.}~[2] addressed link fragility by developing a dynamic topology optimization methodology, where the network must be recomputed faster than links degrade. In virtual circuit networks, Harshavardhana~[3] designed optimal topologies that guarantee loop-free alternate routing while minimizing total link length. In power grids, topology optimization---often called optimal transmission switching---has proven effective for reducing operational costs and managing congestion~[4--6]. In data communication networks, topology optimization determines how switches and routers interconnect servers, directly impacting latency, throughput, and energy consumption; Lebiednik~\textit{et al.}~[7] surveyed data center network topologies such as Fat-tree and BCube, while Xu~\textit{et al.}~[8] examined routing optimization strategies and their dependence on underlying network structure. 

\vspace{-2mm}
\subsection{The Steiner Tree Problem and GeoSteiner Algorithm}

Our work is motivated by network-forming swarm robotics, where mobile robots try to deliver power to a set of terminal nodes while maintaining connectivity to a power source. Connectivity is established by a finite number of mobile robots that form and reconfigure physical links. In such networks, certain robots can serve as relay nodes---intermediate connection points that reduce the total network length required to connect all (or a subset of) the terminals. This naturally leads us to formulate our problem as a variant of the Euclidean Steiner Tree Problem.



The Steiner Tree Problem asks for the shortest network connecting a set of terminal points, allowing the introduction of additional junction points called Steiner points~[9]. Unlike Minimum Spanning Trees (MST), which connect terminals using only direct edges, Steiner trees can reduce total length by introducing intermediate connection points.
The flexibility of Steiner points comes at a cost: finding the optimal Steiner tree is NP-hard~[9]. Several algorithms have been developed to address this complexity. Brazil~\textit{et al.}~[10] developed exact branch-and-bound methods, while Lorenzen and Winter~[11] and Dreyer and Overton~[12] proposed faster heuristics that sacrifice optimality. Extensions such as prize-collecting Steiner trees~[13] incorporate terminal rewards, balancing tree cost against collected prizes. Among these approaches, the GeoSteiner~\textit{et al.}~[14] stands out as the state-of-the-art method for computing provably optimal Steiner trees in the Euclidean plane.

GeoSteiner's framework consists of two stages: (1) generating a set of Full Steiner Trees (FSTs)---locally optimal Steiner trees over subsets of terminals---and (2) aggregating these FSTs via a MILP that selects the minimum-cost combination covering all terminals. This separation of geometric computation from combinatorial optimization makes GeoSteiner particularly well suited as a foundation for our methodology. We inherit the FST generation stage, which provides high-quality geometric building blocks computed in Euclidean space---unlike FLUTE~[15] and other rectilinear Steiner tree algorithms~[16] that use Manhattan distance. However, we replace GeoSteiner's FST aggregation MILP with our own formulation tailored to battery-aware topology optimization, incorporating budget constraints, terminal prioritization, and topology transition costs. By building upon a framework that achieves provably optimal solutions for the classical problem, we expect that pairing its FST generation with our problem-specific MILP will yield high-quality solutions for our extended problem.

\vspace{-3mm}
\subsection{Challenges Beyond the Classic Steiner Tree Problem}

While GeoSteiner solves the classical Steiner Tree Problem optimally, our battery-constrained network problem introduces three key challenges that require algorithmic extensions:
\underline{Dynamic terminal states:} Unlike classical formulations, where terminals have no properties, our terminals have time-varying battery levels. Terminals with low battery require higher priority for connection to prevent depletion. This creates a feedback loop: topology selection determines which terminals charge, and battery states influence which topology is optimal.
\underline{Budget on network length:} In practical deployments, resource limitations prevent connecting all terminals simultaneously. A budget constraint on total network length (i.e., number of available swarm robots) forces partial coverage, requiring the algorithm to select which subset of terminals to connect at each time step. 
\underline{Topology transition costs:} In swarm robotic networks, changing the network topology requires physical relocation of the robots. Frequent topology changes incur significant effort and energy expenditure. Therefore, the algorithm must balance responsiveness to changing battery states against the cost of reconfiguring the network structure.

\vspace{-3mm}
\subsection{Our Contribution}

We present the battery-aware dynamic topology optimization algorithm (BA-DTO) that extends GeoSteiner's Steiner Tree framework to address the challenges outlined above. Our approach formulates Full Steiner Tree (FST) selection as a Mixed-Integer Linear Program (MILP) that incorporates:
\begin{enumerate}[noitemsep, leftmargin=12pt]
    \item \textbf{Budget constraints} limiting total network length, which forces partial terminal coverage and models real-world resource limitations;
    \item \textbf{A battery-aware objective function} that assigns priority weights to terminals based on their battery levels, incentivizing the optimizer to connect low-battery terminals;
    \item \textbf{An overlap-correction term} that prevents double-counting of battery rewards when multiple selected FSTs share terminals, ensuring accurate objective evaluation; and
    \item \textbf{A topology transition cost} that penalizes changes from the previous iteration's configuration, balancing responsiveness to battery dynamics against reconfiguration energy expenditure.
\end{enumerate}

Through iterative re-optimization, our algorithm adapts the network topology as terminal battery states evolve---connected terminals charge while disconnected terminals deplete---maintaining efficient connectivity and preventing any terminal from reaching critical depletion. 

\section{Methodology} \label{Sect:Methodology}

\subsection{Battery-Aware Dynamic Topology Optimization (BA-DTO)}
\begin{wrapfigure}{r}{0.35\textwidth}
    \centering
    \vspace{-0.1cm}
    \includegraphics[width=0.32\textwidth]{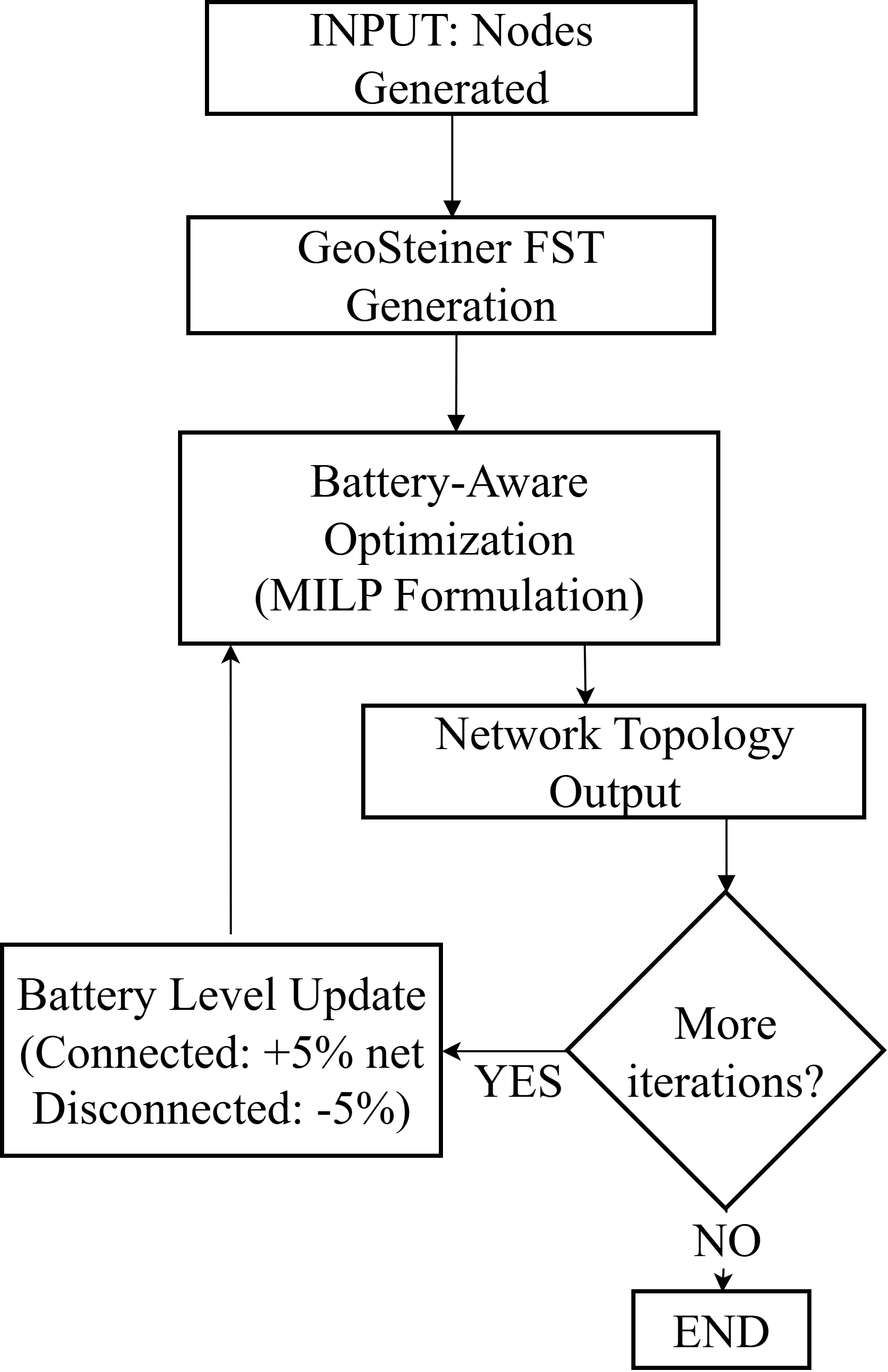}
    \caption{Methodology flowchart.}
    \label{fig:flowchart}
    \vspace{-0.5cm}
\end{wrapfigure}
Our algorithm performs iterative, battery-aware network reconfiguration by solving a novel FST-aggregation problem at each time step. The pipeline (Fig. \ref{fig:flowchart}) begins with terminal coordinates and battery states, generates candidate Full Steiner Trees (FSTs) via GeoSteiner, then selects the optimal subset of FSTs through a Mixed-Integer Linear Program (MILP) that jointly minimizes network length, prioritizes low-battery terminals, respects budget constraints, and penalizes topology transitions. After each iteration, connected terminals charge while disconnected terminals deplete, driving the next optimization cycle. This closed-loop approach enables dynamic topology adaptation as battery states evolve.

\vspace{-3mm}
\subsection{FST Generation via GeoSteiner}

We leverage the GeoSteiner framework for generating candidate FSTs. GeoSteiner takes a set of terminal coordinates $(x,y)$ as input, computes optimal Steiner trees in the Euclidean plane, and generates a set of candidate Full Steiner Trees (FSTs). A Full Steiner Tree (FST) is defined by: (i) a tree that connects a subset of terminals optimally, (ii) may include Steiner points of degree 3, and (iii) it is optimal for its specific terminal subset.
The output from the algorithm is a collection of \textit{m} candidate FSTs. Each FST contains the information on: (i) list of terminals it connects, (ii) positions of Steiner points, (iii) total sum of edge lengths, and (iv) number of edges/vertices. 
To generate the FSTs, the algorithm first enumerates possible terminal subsets. For each subset, it computes an optimal Steiner topology. These computed suboptimal Steiner topologies are immediately pruned. This gives us a comprehensive set covering all the useful terminal combinations. GeoSteiner handles only geometry; our MILP formulation handles budget constraints, battery awareness, and FST selection. Hence, we only use GeoSteiner for FST generation (Geometric computation) and separate geometry from optimization. FSTs' selection with constraints is handled by our MILP formulation and is then optimized.

\subsection{Battery-Aware FST Aggregation}
\begin{wrapfigure}{r}{0.43\textwidth}
\vspace{-0.8cm}
\begin{center}
\fbox{\parbox{0.42\textwidth}{
\small
\setlength{\abovedisplayskip}{3pt}
\setlength{\belowdisplayskip}{3pt}
\begin{align}
\min \quad &\sum_{i=0}^{m-1} \left( c_i + \sum_{j \in T_i} \alpha \left( \frac{b_j}{100} - 1 \right) \right) x_i - \sum_{(i,j) \in \mathcal{M}} D_{ij} y_{ij} \label{eq:obj}\\
\text{s.t.} \quad & \sum_{i=0}^{m-1} c_i x_i \leq B \label{eq:budget}\\
& \sum_{i=0}^{m-1} (|V_i| - 1) x_i + \sum_{j=0}^{n-1} z_j = n - 1 \label{eq:span}\\
& x_i + z_j \leq 1, \; \forall\, i,j: j \in T_i  \label{eq:cov1}\\
& \sum_{i: j \in T_i} x_i + z_j \geq 1, \; \forall\, j \label{eq:cov2}\\
& z_{t_0} = 0 \label{eq:source}\\
& y_{ik} \leq x_i, \; y_{ik} \leq x_k , \; y_{ik} \geq x_i + x_k - 1 \label{eq:ylin}\\
& x_i, z_j, y_{ik} \in \{0,1\} \label{eq:vars}
\end{align}
}}
\end{center}
\vspace{-0.7cm}
\end{wrapfigure}
\textbf{Problem statement:} Given $n$ terminals with 2D coordinates and battery levels $b_k \in [0,100]$, along with $m$ candidate FSTs from GeoSteiner and a normalized budget limit $B$, the problem is to select a subset of FSTs that minimizes network length while prioritizing low-battery terminals. The source terminal $t_0$ must always be connected. The output is a binary decision on which FSTs to select, determining which terminals are covered and defining the network topology for the current iteration. We formulate the FST selection problem as a MILP. Let $x_i \in \{0,1\}$ indicate whether FST $i$ is selected, $z_j \in \{0,1\}$ indicate whether terminal $j$ is uncovered, and $y_{ik} \in \{0,1\}$ be an auxiliary variable for overlap-correction between FSTs $i$ and $k$.

\textbf{Objective Function.}
The objective \eqref{eq:obj} consists of three components. The first component is the construction cost $c_i$, which is the normalized Euclidean length of FST $i$. The second component is the battery-aware term: for each terminal $j$ covered by FST $i$, we add $\alpha(b_j/100 - 1)$, which ranges from $-\alpha$ when $b_j = 0$ (depleted) to $0$ when $b_j = 100$ (fully charged). Since we minimize the objective, FSTs covering low-battery terminals receive more negative costs, making them more attractive to the optimizer. 
The third component is the overlap-correction term, which addresses double-counting when two FSTs share a terminal. The set $\mathcal{M}$ contains all pairs $(i, k)$ of FSTs that share one or more terminals, given by $\mathcal{O}_{ik}$. When both FSTs are selected, the battery reward for the overlapping terminals would be counted twice in the first summation. The correction term $-D_{ik} \cdot y_{ik}$, where $D_{ik} = \sum_{j\in \mathcal{O}_{ik}}\alpha(b_j/100 - 1)$, subtracts the duplicate reward. Since $D_{ik}$ is negative for low-battery terminals, subtracting it adds a positive value that cancels one instance of the double-counted negative reward. This correction term is an artifact of extending the classic Steiner tree formulation with nodal features. The original problem considers only edge lengths, where overlapping FSTs do not cause double-counting.
\textbf{Budget Constraint.}
Constraint \eqref{eq:budget} limits the total network length to budget $B$. This forces partial coverage when the budget is insufficient to connect all terminals, creating the dynamic optimization problem where the algorithm must choose which terminals to prioritize.
\textbf{Spanning Tree Structure.}
Constraint \eqref{eq:span} ensures that the selected FSTs form a valid tree structure. Each FST $i$ with $|V_i|$ vertices contributes $|V_i| - 1$ edges. The constraint requires that the total number of edges in selected FSTs, plus the number of uncovered terminals, equals $n-1$. This relationship holds because: if $n_c$ terminals are covered, the spanning tree over them requires $n_c - 1$ edges, and the remaining $n - n_c$ terminals are uncovered, giving $(n_c - 1) + (n - n_c) = n - 1$.
\textbf{Terminal Coverage.}
Constraints \eqref{eq:cov1} and \eqref{eq:cov2} work together to ensure consistent terminal coverage. Constraint \eqref{eq:cov1} enforces that if any FST covering terminal $j$ is selected ($x_i = 1$), then $j$ must be marked as covered ($z_j = 0$). Constraint \eqref{eq:cov2} enforces that each terminal must either be covered by at least one selected FST or be marked as uncovered ($z_j = 1$).
Together, these constraints establish a one-to-one correspondence: $z_j = 0$ if and only if at least one FST covering $j$ is selected. In practice, we implement a continuous relaxation on $z_j \in [0,1]$ (rather than binary), which allows computational tractability, though the constraints naturally drive $z_j$ to binary values at optimality.
\textbf{Source Connectivity.}
Constraint \eqref{eq:source} requires the source terminal $t_0$ (the charging station) to always be covered, ensuring the network remains connected to the power source.
\textbf{Linearization of Overlap-Correction.}
Constraint \eqref{eq:ylin} linearizes the product $y_{ik} = x_i \cdot x_k$ using the standard McCormick envelope. The three inequalities ensure that $y_{ik} = 1$ if and only if both $x_i = 1$ and $x_k = 1$, enabling the overlap-correction to apply precisely when both overlapping FSTs are selected.

\vspace{-3mm}
\subsection{Topology Transition Cost}

While dynamic topology adaptation is crucial for maintaining battery levels across the network, frequent reconfiguration incurs costs. In physical networks such as swarm robotic networks, each topology change consumes energy for repositioning and link establishment. Zhao \textit{et al.}~[17] addressed similar reconfiguration costs in data center networks by introducing switching penalties into their optimization formulation. To balance battery-aware adaptation with network stability, we extend the objective with a graph distance penalty,
$
\gamma \sum_{i=0}^{m-1} |x_i - x_i^{(t-1)}|,
$
where $x_i^{(t-1)}$ is the FST selection from the previous iteration and $\gamma$ is the penalty weight. This term penalizes topology changes, reducing unnecessary network switching while still allowing adaptation when battery states demand it.

\vspace{-3mm}
\subsection{Implementation and Complexity}

The MILP is solved using IBM ILOG CPLEX with 1\% relative MIP gap tolerance. Computational complexity is dominated by $O(m^2)$ auxiliary variables for overlap-correction linearization. For 20 terminals with 40 FSTs, 269 variables and 791 constraints were generated; 30 iterations solved in 28 seconds ($\approx$0.93 seconds/iteration) on a Dell XPS 9320 laptop (Intel i7-1360P, 32 GB RAM). For larger networks (30+ terminals), runtime increases quadratically due to growth in overlapping FST pairs.

\section{Numerical Simulations} \label{sec:simulations}

\subsection{Experimental Setup}

We simulated 20 terminals that are randomly distributed in a unit square $[0,1] \times [0,1]$ with uniformly generated coordinates; typical distances between neighboring terminals are approximately 0.1--0.2. The budget constraint $B = 1.6$ is the normalized value where each FST's cost is divided by the maximum FST cost in the current iteration.This budget value results in a partial coverage of 60-75\% of terminals, creating a dynamic switching behavior. In practice, $B$ represents the total relay capacity available; for example, the number of robots that can form connections divided by the maximum network span.  The battery weight $\alpha = 10.0$ prioritizes low-battery terminals while under the budget constraint. Connected terminals charge at 10\% per iteration, and all terminals discharge at 5\%, giving a net gain of 5\% if connected. Finally, the graph distance penalty weight $\gamma = 1.0$ balances topology stability and battery-aware changes.

\vspace{-2mm}
\subsection{Numerical Results}

\begin{figure}[htb]
    \centering
    \begin{subfigure}[b]{0.24\textwidth}
        \centering
        \includegraphics[width=\textwidth, trim=50 120 0 0, clip]{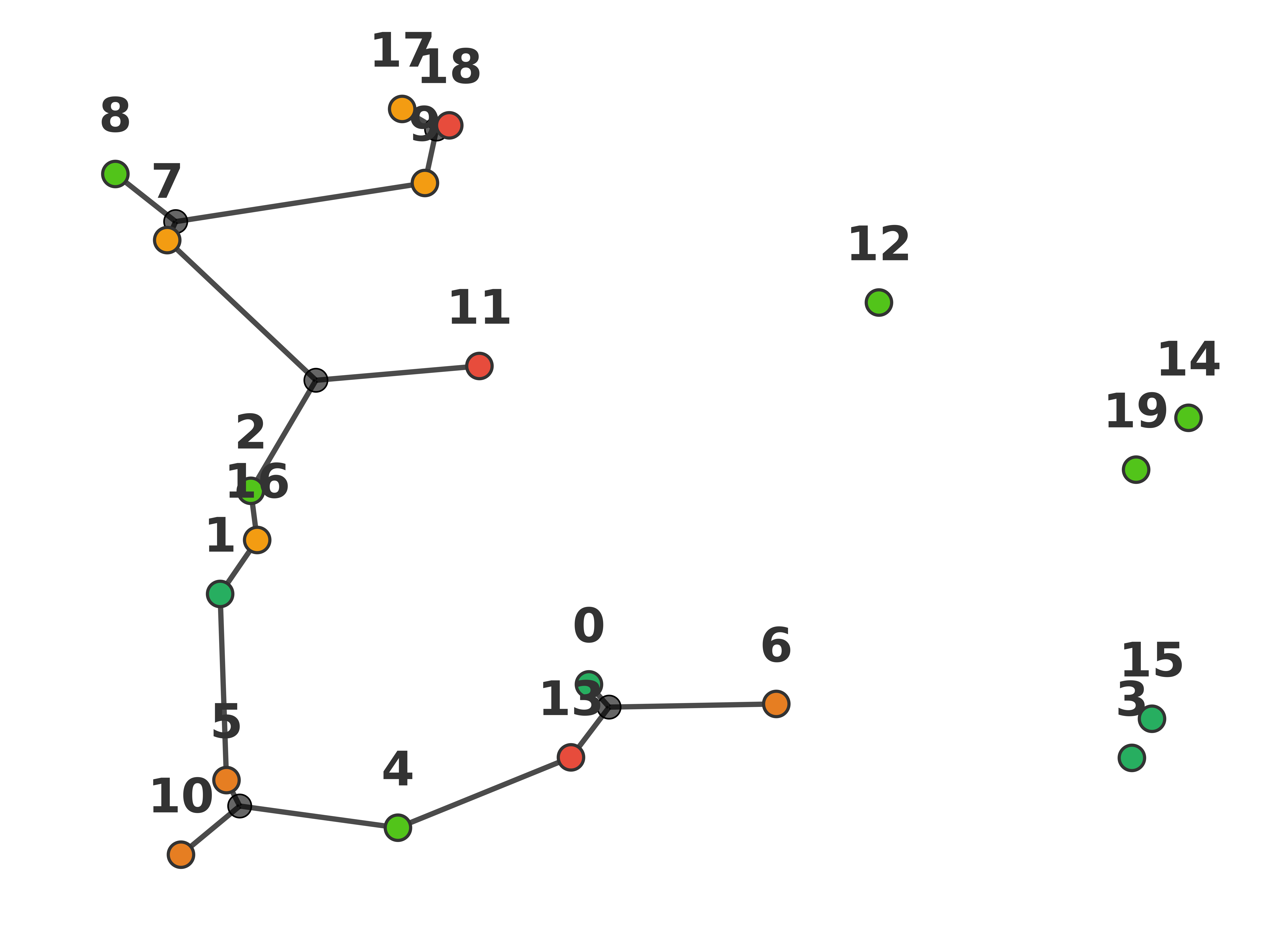}
        \caption{Iteration 1}
        \label{fig:topology_iter1}
    \end{subfigure}\hfill
    \begin{subfigure}[b]{0.24\textwidth}
        \centering
        \includegraphics[width=\textwidth, trim= 50 120 0 0, clip]{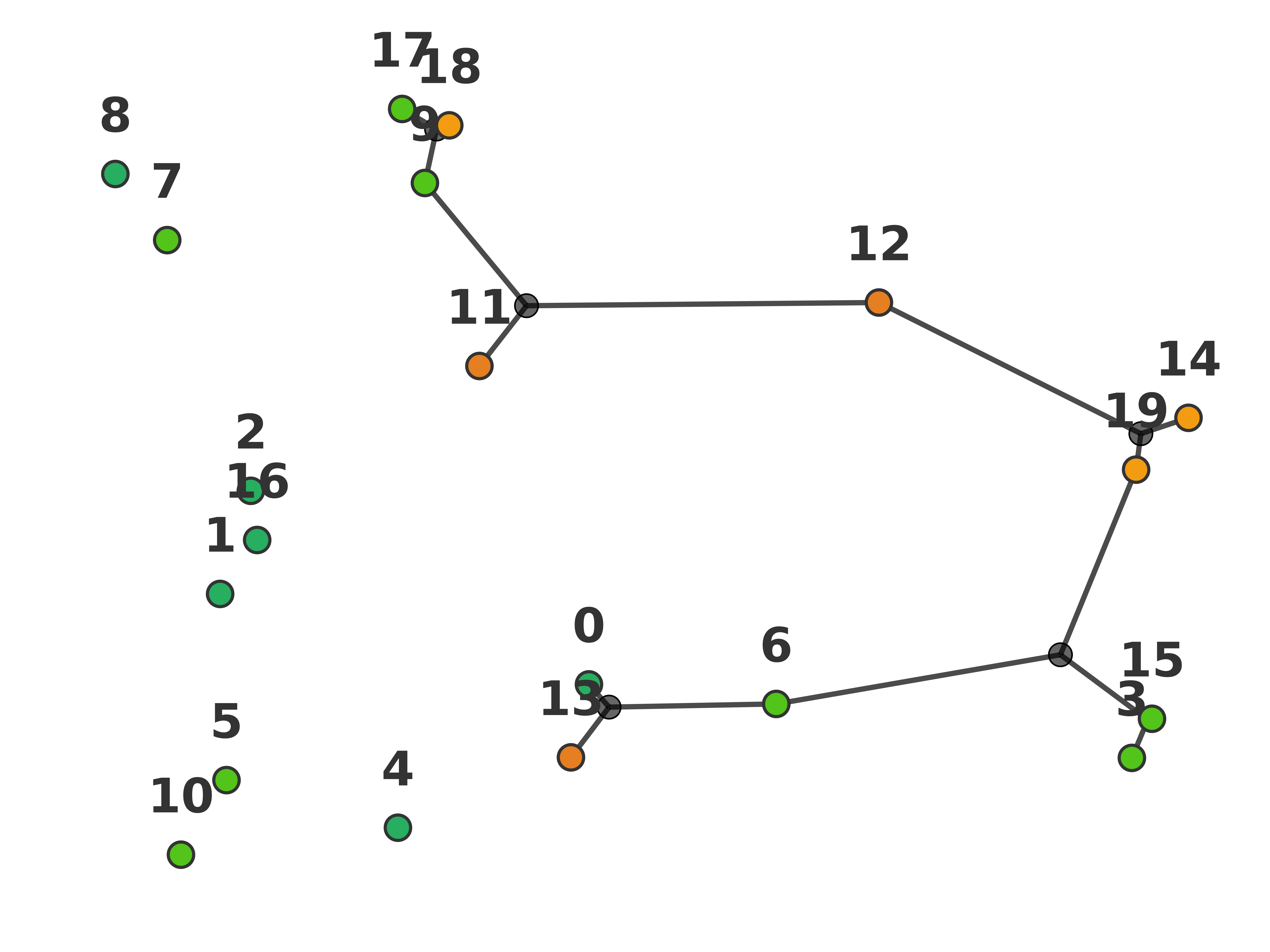}
        \caption{Iteration 7}
        \label{fig:topology_iter7}
    \end{subfigure}\hfill
    \begin{subfigure}[b]{0.24\textwidth}
        \centering
        \includegraphics[width=\textwidth, trim= 50 120 0 0, clip]{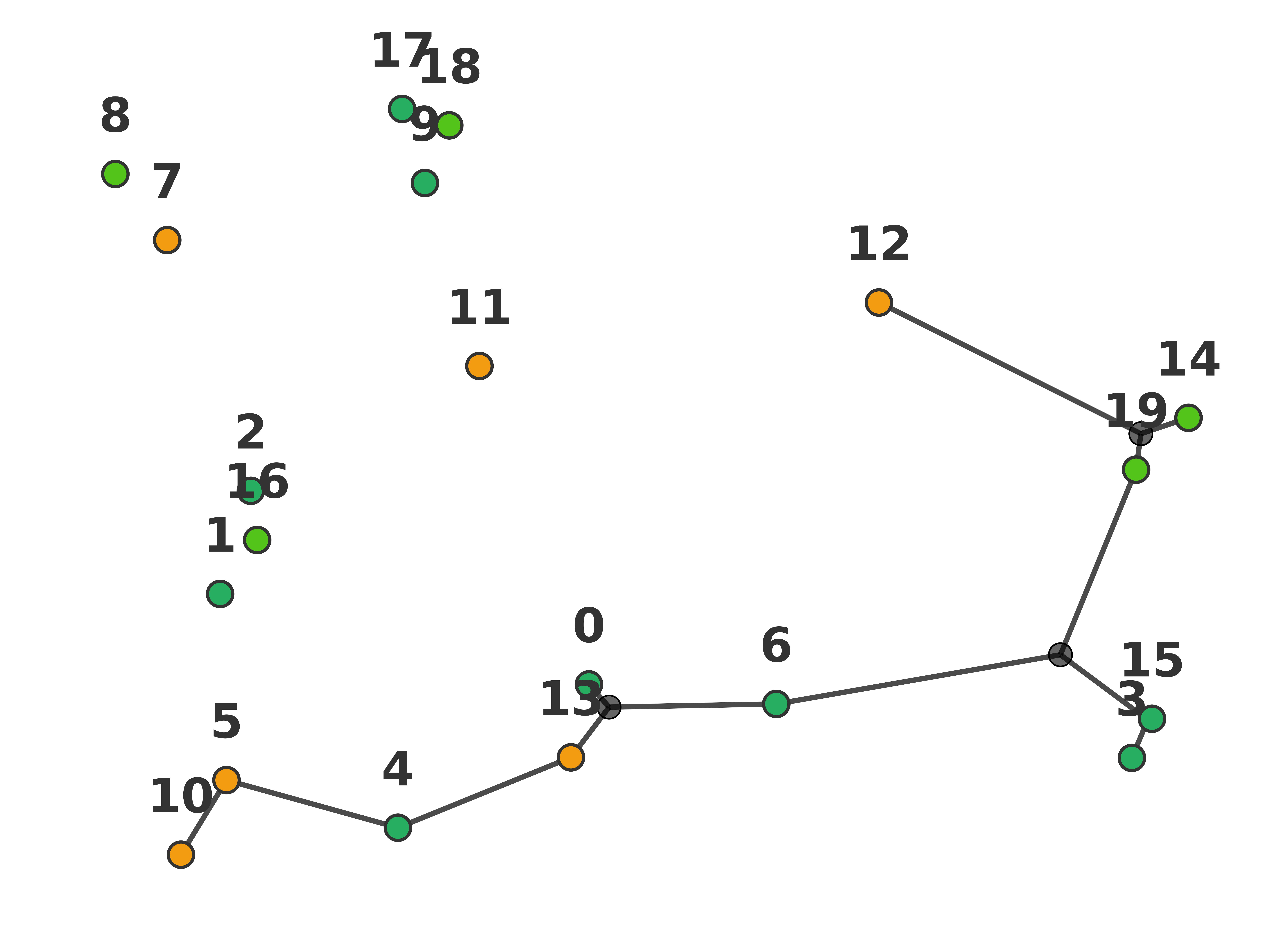}
        \caption{Iteration 11}
        \label{fig:topology_iter11}
    \end{subfigure}\hfill
    \begin{subfigure}[b]{0.24\textwidth}
        \centering
        \includegraphics[width=\textwidth, trim= 50 120 0 0, clip]{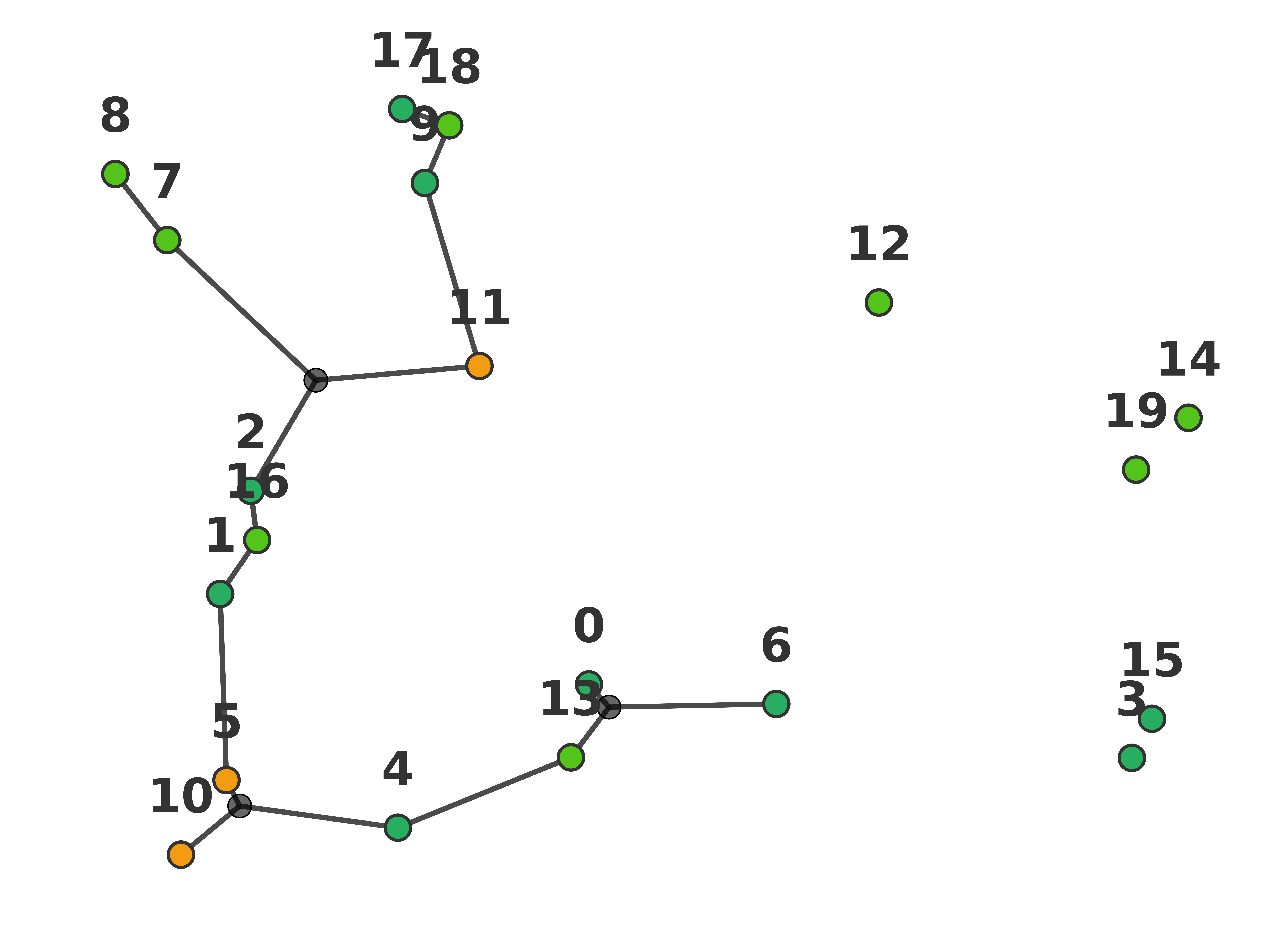}
        \caption{Iteration 12}
        \label{fig:topology_iter12}
    \end{subfigure}
    
    \vspace{0.5em}
    
    \begin{subfigure}[b]{0.24\textwidth}
        \centering
        \includegraphics[width=\textwidth, trim= 50 120 0 0, clip]{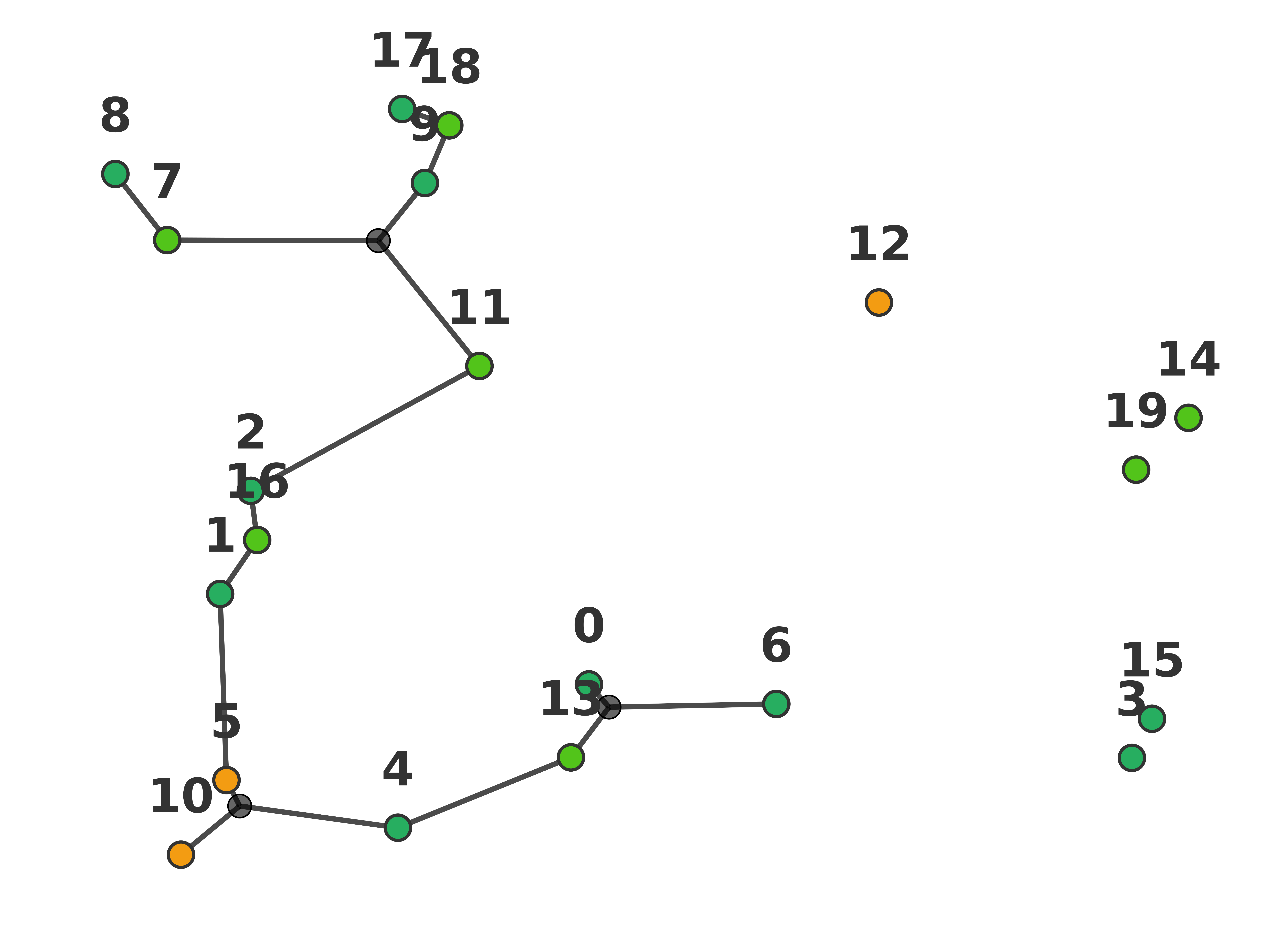}
        \caption{Iteration 13}
        \label{fig:topology_iter13}
    \end{subfigure}\hfill
    \begin{subfigure}[b]{0.24\textwidth}
        \centering
        \includegraphics[width=\textwidth, trim= 50 120 0 0, clip]{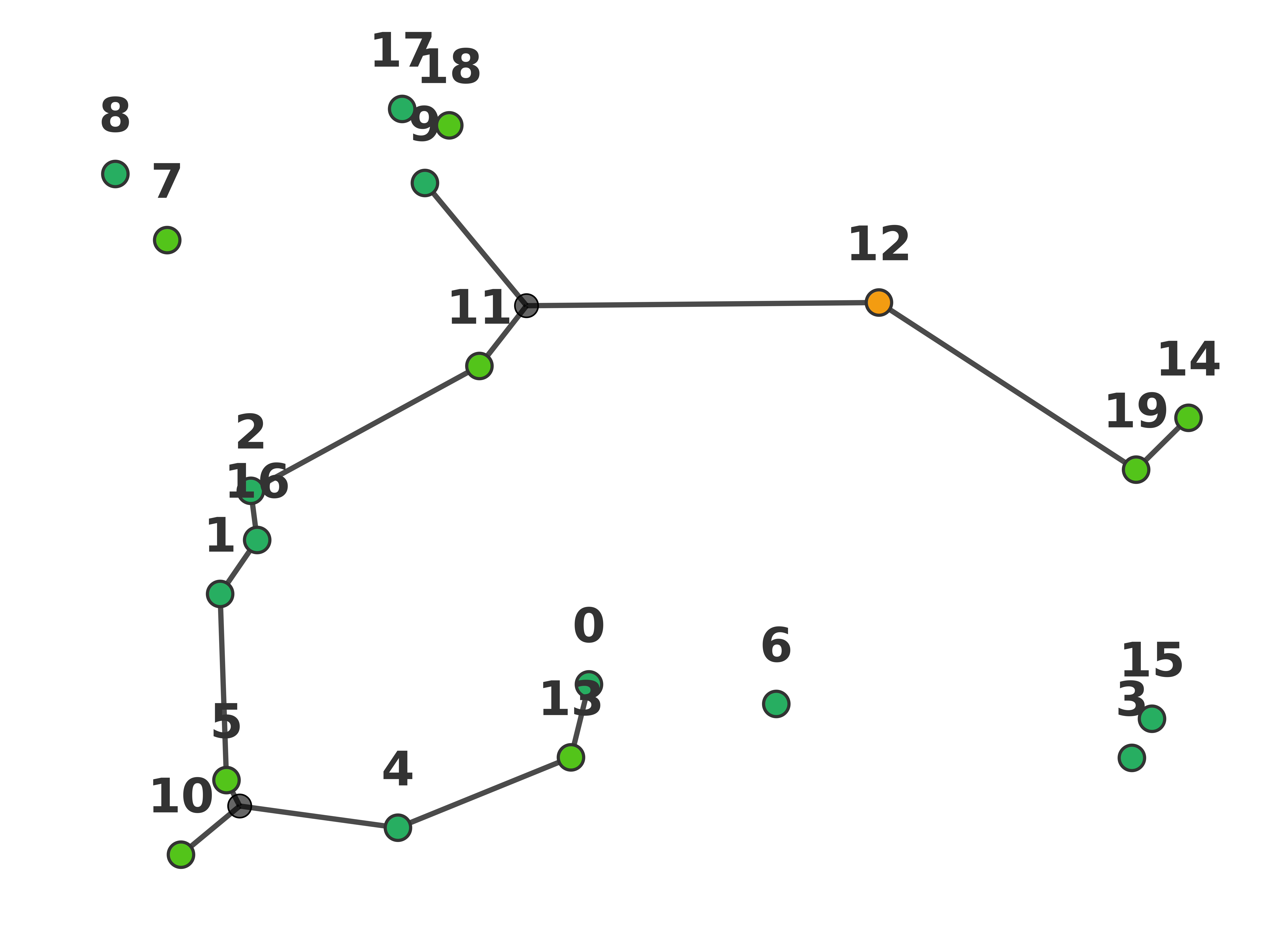}
        \caption{Iteration 15}
        \label{fig:topology_iter15}
    \end{subfigure}\hfill
    \begin{subfigure}[b]{0.24\textwidth}
        \centering
        \includegraphics[width=\textwidth, trim= 50 120 0 0, clip]{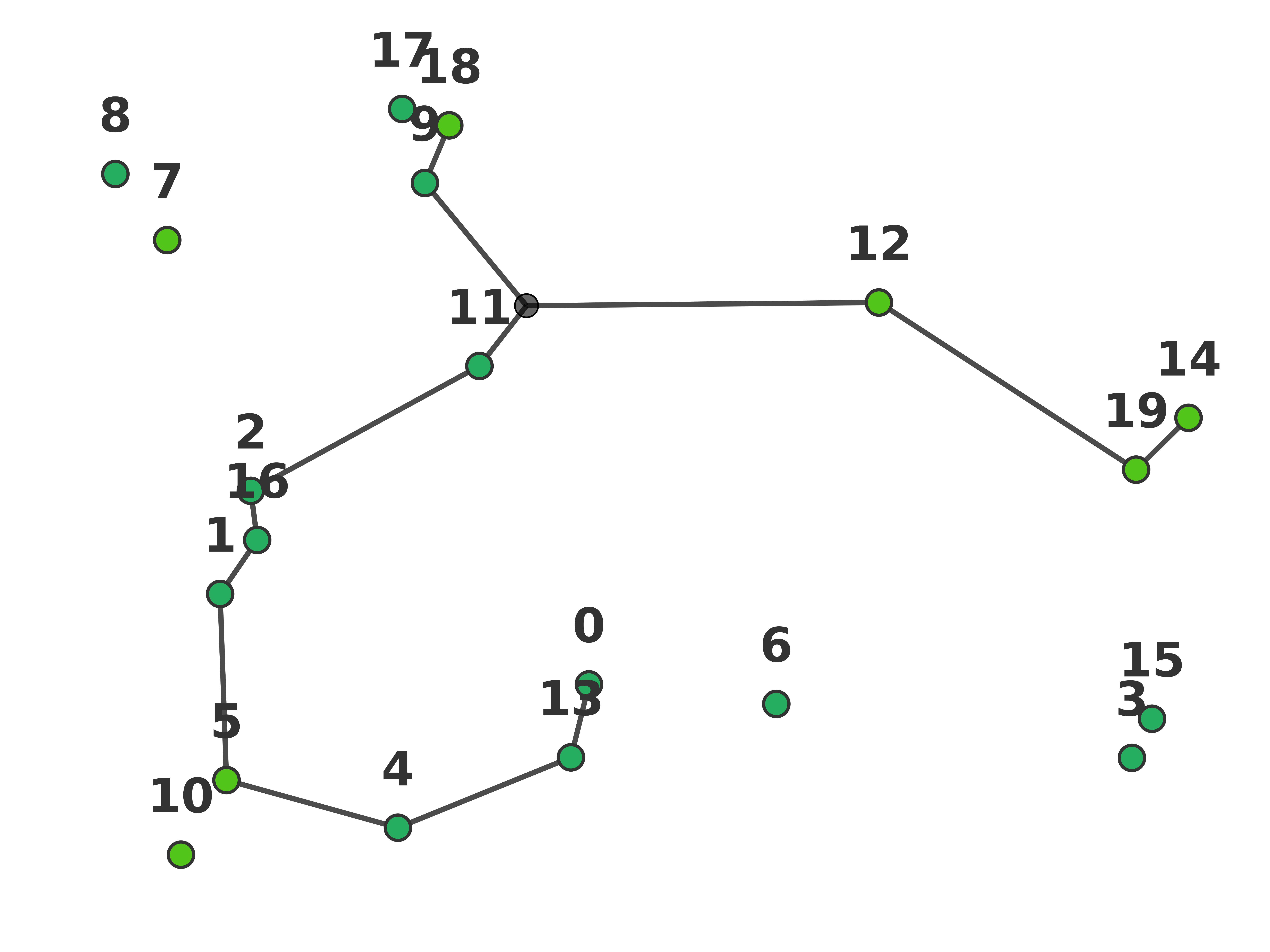}
        \caption{Iteration 17}
        \label{fig:topology_iter17}
    \end{subfigure}\hfill
    \begin{subfigure}[b]{0.24\textwidth}
        \centering
        \includegraphics[width=\textwidth, trim= 50 120 0 0, clip]{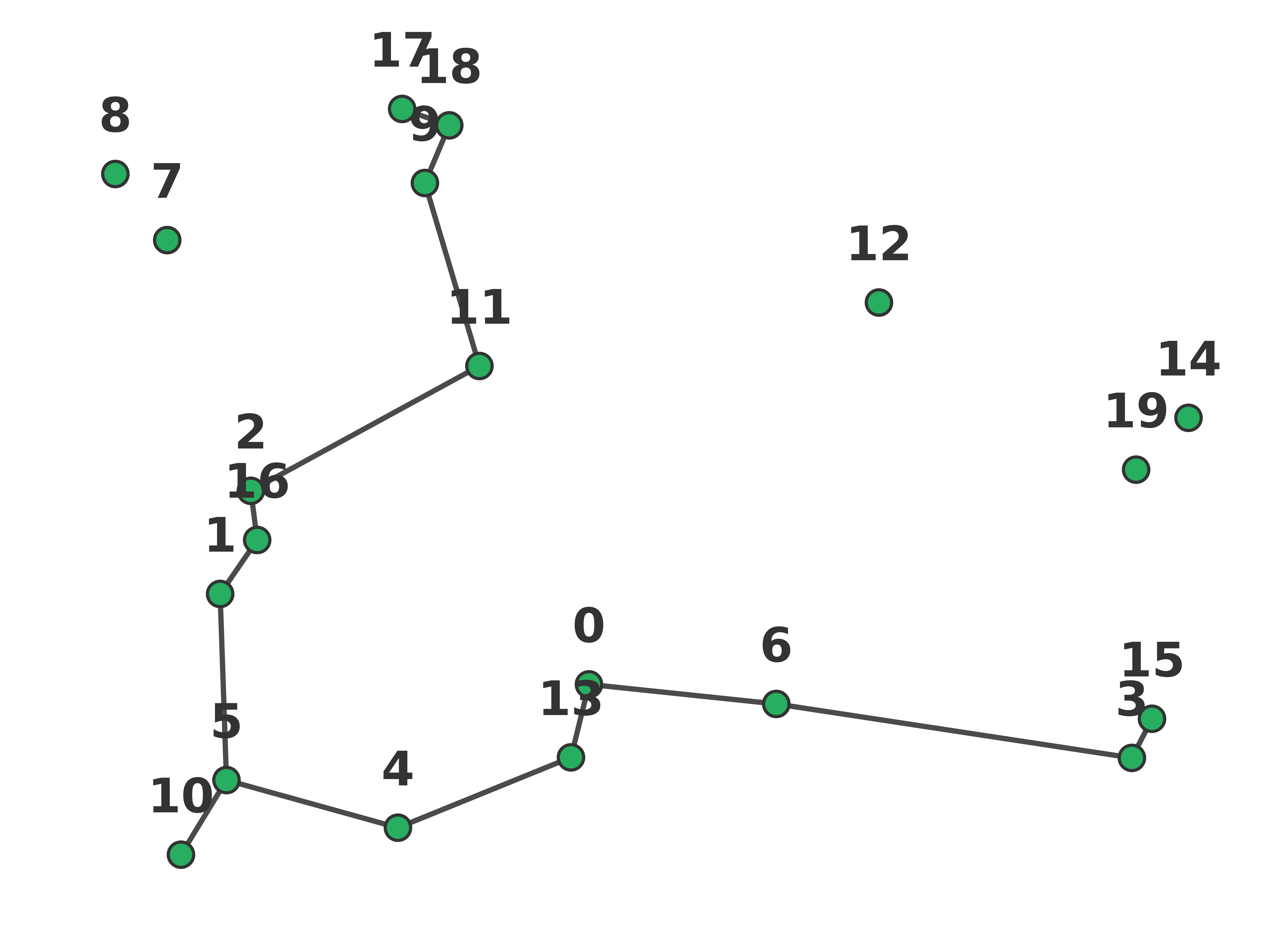}
        \caption{Iteration 27}
        \label{fig:topology_iter27}
    \end{subfigure}

    \caption{Topology evolution across iterations. Terminal colors indicate battery levels. Black dots are Steiner points.}
\end{figure}

Figures \ref{fig:topology_iter1}-\ref{fig:topology_iter27} show the network topology evolution across iterations. Terminal colors indicate the battery levels (green = high, red/orange = low). The black dots represent Steiner points and the terminal indexed 0 is the power source. In iteration 1, terminals have heterogeneous battery levels, so the optimizer selects FSTs that have low-battery levels while satisfying the budget constraint. By iteration 7, previously connected terminals show improved battery levels while disconnected terminals have depleted. Hence, the topology switches to account for the increased priority of the depleted terminals, as seen in Figures \ref{fig:topology_iter11}-\ref{fig:topology_iter13}. The topology reconfiguration occurs only when the priority level of the disconnected terminals become higher than connected terminals. By iteration 27, all terminals are close to fully charged.
\begin{wraptable}{r}{0.6\textwidth}
    \vspace{-3mm}
    \centering
    \small
    
    \captionof{table}{Performance Metrics Across Iterations}
    \label{tab:metrics}
    \begin{tabular*}{\linewidth}{@{\extracolsep{\fill}}lccc}
    \hline
    \textbf{Metric} & \textbf{Iter 1} & \textbf{Iter 15} & \textbf{Iter 30} \\
    \hline
    Terminals covered & 15 / 20 (75\%) & 14 / 20 (70\%) & 14 / 20 (70\%) \\
    Min Battery (\%) & 2.7 & 49.0 & 68.6 \\
    Avg Battery (\%) & 56.8 & 79.3 & 89.5\\
    Tree Length (norm.) & 1.475 & 1.576 & 1.556 \\
    Budget Utilization & 92.2\% & 98.5\% & 97.2\% \\
    \hline
    \end{tabular*}
        
    \captionof{table}{Comparison of Methods}
    \label{tab:baseline_comparison}
    \begin{tabular*}{\linewidth}{@{\extracolsep{\fill}}lccc}
    \hline
    \textbf{Method} & \textbf{Initial Min} & \textbf{$\Delta$Min Battery} & \textbf{Total Edge Changes} \\
    \hline
    MST & 2.7\% & +2.3\% & 327 \\
    BA-DTO-1  & 2.7\% & +68.6\% & 407 \\
    BA-DTO-2  & 2.7\% & +65.9\% & 113\\
    \hline
    \end{tabular*}
    \vspace{-3mm}
\end{wraptable}
Table \ref{tab:metrics} shows that the algorithm successfully prevents terminal depletion. The minimum battery increases from 2.7\% to 68.6\%. The average battery level across all the terminals is increased from 56.8\% to 89.5\%, and the coverage decreases from 75\% to 70\%, meaning there are fewer critical terminals and the optimizer can leave more high-battery terminals disconnected. The budget utilization remains above 92\%, showing efficient resource use.

\vspace{-3mm}
\subsection{Parameter Sensitivity}

The budget constraint $B$ controls the coverage-stability trade-off. Through experimentation, we found $B = 1.6$ provides a balance with 60--70\% coverage. Lower values ($B < 1.4$) reduce coverage below 50\% and increase computation time, while higher values ($B > 1.8$) result in full coverage, violating our partial connectivity assumption.
The battery weight $\alpha$ controls priority given to low-battery terminals. At $\alpha = 5$, the network prioritizes length over battery, leaving low-battery terminals unreached. At $\alpha = 10$, the optimizer balances connecting low-battery terminals while maintaining distant connections. Values below 5 violate budget or coverage constraints, while values above 10 result in minimal topology switching as battery reward dominates length cost.
The graph distance weight $\gamma = 1$ controls network stability penalty. Higher values prevent topology changes entirely, while lower values promote frequent switching that improves battery balance but increases energy consumption. At $\gamma = 1$, the algorithm allows topology adaptation when unconnected terminals have higher priority due to battery depletion.

\vspace{-3mm}
\subsection{Performance Evaluation and Comparison}

To evaluate the effectiveness of our battery-aware formulation and graph distance penalty, we compare three approaches using identical conditions. MST-only connects all terminals using only terminal-to-terminal edges without Steiner points. It selects the minimum spanning tree that satisfies the budget constraint. BA-DTO-1 uses our battery-aware FST selection with Steiner points but without the graph distance penalty ($\gamma = 0$). BA-DTO-2 adds the graph distance penalty ($\gamma = 1.0$) to minimize the topology reconfiguration between iterations.


Table \ref{tab:baseline_comparison} shows the results after 30 iterations. The minimum battery level of the MST-only approach only improves by 2.3\%, while BA-DTO-1 and BA-DTO-2 have an improvement of 68.6\% and 65.9\%. This demonstrates a notable advantage in using Steiner points for the topology. 
Comparing BA-DTO-1 and BA-DTO-2, both have a similar amount of improvement in the battery level; however, BA-DTO-2 reduces the total topology reconfiguration from 407 edges to 113 edges, which is a 72.2\% reduction. This is a significant improvement in network stability with battery-awareness, and a notable reduction in energy spent on network reconfiguration.
These results highlight two key findings. First, Steiner points provide essential geometric flexibility that MST-only approaches lack. This allows the optimizer to route through intermediate points to connect low-battery terminals more effectively regardless of their spatial distribution. Second, the graph distance penalty achieves substantial stability improvements with minimal impact on battery percentage; only a 2.7\% battery improvement difference between BA-DTO-1 and BA-DTO-2 while achieving a 72.2\% reduction in edge changes. This demonstrates that network stability and battery-awareness can be optimized together without significant trade-offs, making the approach practical for real-world deployments.

\section{Conclusion} \label{sec:conclusion}

This paper introduced BA-DTO, a battery-aware topology optimization framework built on GeoSteiner's Full Steiner Tree (FST) generation model. By formulating FST selection as a MILP that minimizes the length cost and prioritizes low-battery terminals, the method enables a dynamic, energy-aware network formation. An overlap-correction term was included to accurately represent terminal features when multiple FSTs share terminals, ensuring valid and efficient network structures. 


\end{document}